\documentclass{blois}
\usepackage{float}

\def\be{\begin{equation}}
\def\ee{\end{equation}}
\def\bea{\begin{eqnarray}}
\def\eea{\end{eqnarray}}

\newcommand{\Photo}{}

\begin{document}
\vspace*{4cm}
\title{Top Quark Properties at the LHC}

\author{ K. Beernaert on behalf of the ATLAS and CMS collaborations }

\address{Universiteit Gent, Department of Physics, \\
Proeftuinstraat 86, 9000 Gent, Belgium}

\maketitle\abstracts{The presented results on top quark properties include measurements of the top 
quark pair 
charge asymmetry, the top quark polarization in pair production and single top production, the W helicity in top quark decays. 
Results of spin correlation in top quark pairs are presented and interpreted in terms of the Standard Model (SM) predicted 
values. The cross section of $t\overline{t}$ events produced in association with a W, Z boson or a photon is also presented. 
The results are compared with predictions from the standard model as well as new physics models.} 

\section{Introduction}
\label{sec:introduction}
The top quark decays before it can form bound states and before their spins decorrelate. As a consequence we can study "bare" quark properties. The 
top quark has a mass of $\approx 173$ GeV and is still the heaviest particle known to date and may play a significant role in 
electro-weak symmetry breaking due to its large coupling to the Higgs boson. Measurements of the top quark properties with 
increasing levels of precision test the SM and have the possibility to probe new physics. The data were collected in 
pp collisions in 2011 and 2012 at center-of-mass energies $\sqrt{s}$ of 7 and 8 TeV at the Large Hadron Collider (LHC) by the 
ATLAS~\cite{1748-0221-3-08-S08003} and CMS~\cite{Chatrchyan:2008aa} detectors.

\section{Production Asymmetries}
\label{sec:Production Asymmetries}
Top quark asymmetries were first observed at the Tevatron in the form of a forward-backward asymmetry $A_{FB}$. At the LHC, 
top quark charge asymmetries $A_{C}$ are evaluated by using the decay leptons or by fully reconstructing the top 
quark pair. The charge asymmetry is given by 

\begin{equation}
 A^{ll}_{C} = \frac{N(\Delta \left| \eta_{l} \right| > 0) - N(\Delta \left| \eta_{l} \right| < 0)}{N(\Delta \left| \eta_{l} \right| > 0) + N(\Delta \left| \eta_{l} \right| < 0)}
\label{eq.ChargeAsymLep}
 \end{equation}
\noindent
where $\Delta \left| \eta_{l} \right|\,=\,\left|\eta_{l^{+}}\right| - \left|\eta_{l^{-}} \right|$ with $\eta$ the pseudo-rapidity for the leptonic asymmetry 
and given by 

\begin{equation}
  A_{C} = \frac{N(\Delta \left| y \right| > 0) - N(\Delta \left| y \right| < 0)}{N(\Delta \left| y \right| > 0) + N(\Delta \left| y \right| < 0)}
\label{eq.ChargeAsym}
\end{equation}
\noindent
where $\Delta \left| y \right|\,=\, \left| y_{t} \right| - \left| y_{\overline{t}} \right|$ with $y$ the rapidity. In the SM, the top quark production 
asymmetries are due to NLO QCD interference effects. Measurements of the $A_{FB}$ at the Tevatron and $A_{C}$ at the LHC are 
complementary to evaluate new physics models~\cite{PhysRevD.84.115013,AguilarSaavedra:2011ug}.\\ Two analyses are presented 
in the lepton+jets 
channel. The ATLAS result~\cite{Aad:2013cea} uses 4.7 fb$^{-1}$ of data at $\sqrt{s}\,=$ 7 TeV. 
The CMS result~\cite{CMS-PAS-TOP-12-033} is evaluated at $\sqrt{s}\,=$ 8 TeV using 19.7 fb$^{-1}$ of data. In both analyses 
the top 
quark pairs are reconstructed using a kinematic fit and the kinematic distributions of $\Delta\left|y\right|$ are unfolded 
to parton level. An inclusive result for the charge asymmetry was obtained of $A_{C}\,=\,0.006\,\pm\,0.010\,\mathrm{(stat+syst)}$
by ATLAS and $A_{C}\,=\,0.005\,\pm\,0.007\,\mathrm{(stat)}\,\pm\,0.006\,\mathrm{(syst)}$ by CMS to be compared to the SM 
value~\cite{BernreutherAsym} of 
$A_{C}\,=\,0.0123\,\pm\,0.0005$. Differential measurements of the charge asymmetry are presented as a function of a variety of 
variables such as the invariant mass $m(t\overline{t})$, transverse momentum $p_{T}(t\overline{t})$ and rapidity 
$\left| y(t\overline{t}) \right|$ of the top quark pair. The comparison of the differential measurements with a 
Beyond-Standard-Model (BSM) model that includes axial-vector couplings is made. 
Axi-gluon models with scales of about 1.5 TeV cannot be excluded yet.
A measurement of the charge asymmetry in the dilepton channel is presented by ATLAS~\cite{ATLAS2015} at $\sqrt{s}\,=$ 7 TeV 
using 
4.6 fb$^{-1}$ of data. The $t\overline{t}$ reconstruction is performed using the neutrino weighting technique. The combined 
result of the $ee$, $e\mu$ and $\mu\mu$ channels gives 
$A^{ll}_{C}\,=\,0.024\,\pm\,0.015\,\mathrm{(stat)}\,\pm\,0.009\,\,\mathrm{(syst)}$ based on the decay leptons and 
$A_{C}\,=\,0.021\,\pm\,0.025\,\mathrm{(stat)}\,\pm\,0.017\,\mathrm{(syst)}$ based on the reconstructed top quarks. 

\section{Polarization}
\label{sec:Polarization}
In top quark pair production, a significant polarization can only be due to new physics effects. In a measurement of the top 
quark polarization by the 
ATLAS collaboration~\cite{PhysRevLett.111.232002}, the polarization is assumed to be due to either CP conserving (CPC) or CP 
violating (CPV) new physics processes. In this analysis, the polarization is measured in the lepton+jets and dilepton channel 
at $\sqrt{s}\,=\,7\,\mathrm{TeV}$, giving $\alpha_{l}P_{CPC}\,=\,-0.035\,\pm\,0.014\,\mathrm{(stat)}\,\pm\,0.037\,\mathrm{(syst)}$ 
for the CP conserving process and $\alpha_{l}P_{CPV}\,=\,0.020\,\pm\,0.016\,\mathrm{(stat)}\,{}^{+0.013}_{-0.017}\,\mathrm{(syst)}$ 
for the CP violating process where $\alpha_{l}$ is the spin analyzing power. The results are consistent with the CMS 
result~\cite{PhysRevLett.112.182001}. In single top production, top quarks are expected to 
be almost 100 $\%$ polarized. CMS has presented a 
measurement of the polarization in single top production at $\sqrt{s}\,=$ 8 TeV using 20 fb$^{-1}$ of data. 
In a combination of the electron and muon channel, a 
$P\,=\,0.82\,\pm\,0.12\,\mathrm{(stat)}\,\pm\,0.32\,\mathrm{(syst)}$ was observed~\cite{CMS-PAS-TOP-13-001}.

\section{Spin Correlations}
\label{sec:Spin Correlations}
\begin{figure}[ht!]
 \begin{minipage}{0.5\textwidth}
 \includegraphics[width=\textwidth]{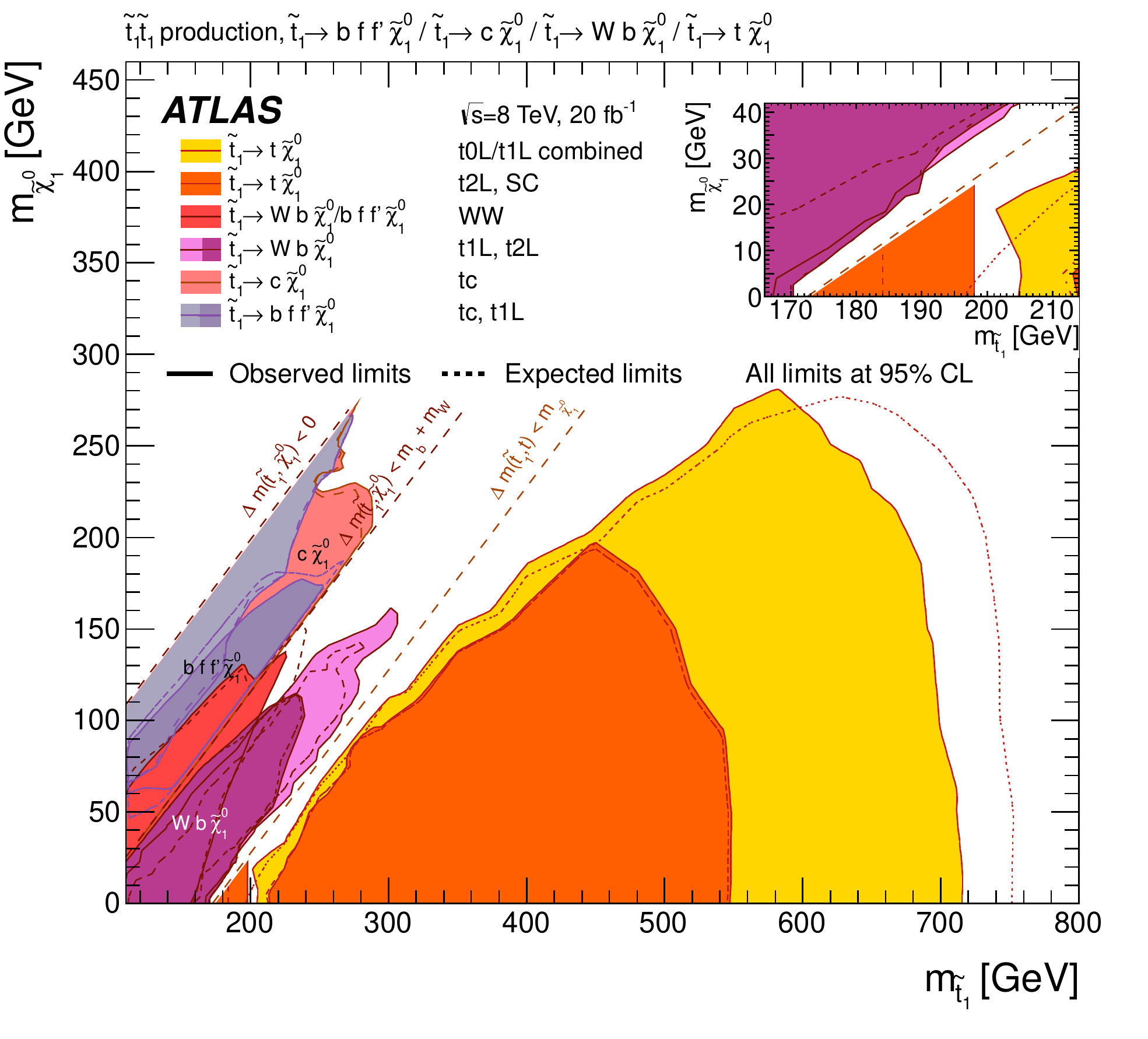}
  \caption{Summary of the dedicated ATLAS searches for top squark pair production at $\sqrt{s}\,=$ 8 TeV. Exclusion limits are 
  shown in the stop1-neutralino1 mass plane.~\protect\cite{Aad:2015pfx}}
\label{fig.ATLASSpinCor}
  \end{minipage}
\begin{minipage}{0.5\textwidth}
 \includegraphics[width=\textwidth]{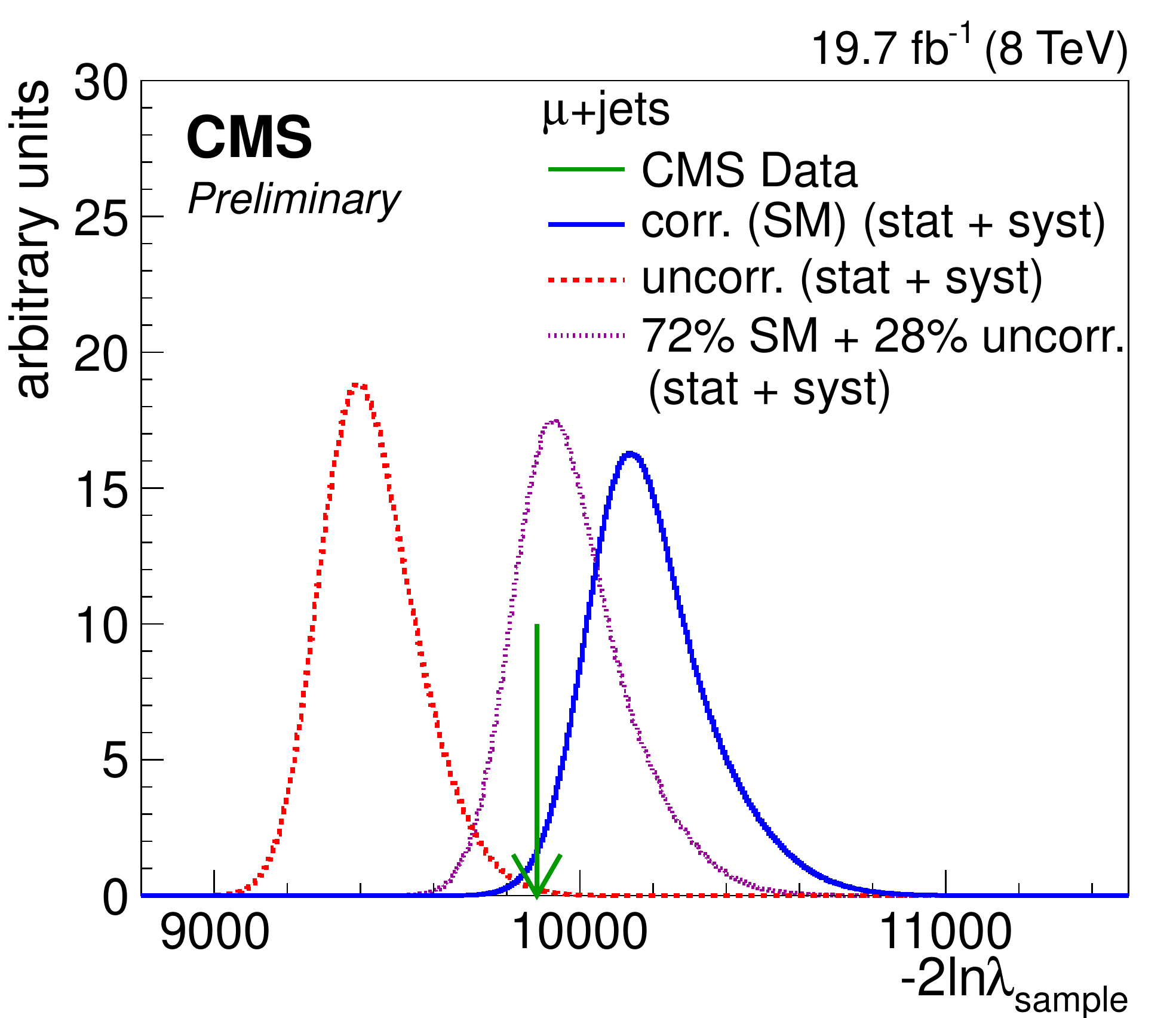}
 \caption{The $-2ln\lambda_\mathrm{sample}$ distribution is shown in simulation, evaluated at the dataset size with the 
 green arrow indicating data. The magenta dotted curve shows an admixture of SM correlated and uncorrelated 
 $t\overline{t}$ events, as observed in the template fit.~\protect\cite{CMS-PAS-TOP-13-015}}
 \label{fig.CMSSpinCor}
\end{minipage}

\end{figure}

In the SM, spin correlation within a top quark pair is predicted. The spin correlation strength has been measured 
by CMS in the dilepton channel~\cite{PhysRevLett.112.182001} at $\sqrt{s}\,=$ 7 TeV using 5.0 fb$^{-1}$ of data. 
In this analysis, the opening angle distribution between the decay leptons has 
been unfolded to parton level to show agreement with the SM up to next-to-leading order (NLO) QCD corrections. In an 
extension of 
this analysis, the differential cross sections have been used to search for top chromomagnetic 
couplings~\cite{CMS-PAS-TOP-14-005}. The real part of the chromomagnetic dipole moment Re$(\mu_{t})$ was measured to be 
$0.037\,\pm\,0.041\,\mathrm{(tot.)}$. Limits on Re$(\mu_{t})$ were set of $-0.043\,<\,\mathrm{Re}(\mu_{t})\,<\,0.117$ at 
95 \% C.L. The result is limited by the scale uncertainties on the theory predictions.\\
In an analysis in the dilepton channel at $\sqrt{s}\,=$ 8 TeV using an integrated luminosity of 20.3 fb$^{-1}$ by 
the ATLAS collaboration~\cite{PhysRevLett.114.142001}, the fraction $f^{SM}$ of top quark pairs showing SM spin correlation 
strength is 
extracted as $f^{SM}\,=\,1.20\,\pm\,0.05\,\mathrm{(stat)}\,\pm\,0.13\,\mathrm{(syst)}$ by using a template fit to the 
opening angle distributions of the decay products. The results have been interpreted in a Minimal Super-Symmetric Model (MSSM) 
where stop squarks decay 100 \% into a top quark and a neutralino with the stop squark mass very close to the top quark mass. 
This region of the 
MSSM phase space is very difficult to access using traditional search techniques~\cite{Aad:2015pfx}. Stop squark masses 
between the top quark mass 
and 191 GeV have been excluded with 95 \% C.L. as shown in fig.~\ref{fig.ATLASSpinCor}. In an analysis 
presented by the CMS collaboration~\cite{CMS-PAS-TOP-13-015} at $\sqrt{s}\,=$ 8 TeV and using an 
integrated luminosity of 19.7 fb$^{-1}$, the fraction $f^{SM}$ of SM top quark pairs has been measured as 
$f^{SM}\,=\,0.72\,\pm\,0.09\,\mathrm{(stat)}\,{}^{+0.15}_{-0.13}\,\mathrm{(syst)}$ by performing a template fit. By using 
a matrix element method, the event likelihoods were calculated under the SM and uncorrelated $t\overline{t}$ hypothesis to 
obtain a variable sensitive to the $t\overline{t}$ spin correlations. In addition, a hypothesis testing procedure was 
performed, shown in fig.~\ref{fig.CMSSpinCor}. 

\section{W Helicity}
\label{sec:W Helicity}
In the SM, the top quark decays almost exclusively to a W boson and a b quark. The W helicity fractions in the decay are 
predicted to be about 70 \% longitudinal $F_0$, 30 \% left-handed $F_L$ and a negligible right-handed 
contribution $F_R$. These helicity fractions influence the top quark partial width and can probe the anomalous couplings. A 
measurement of the W helicity fraction in single top production is presented by the CMS 
collaboration~\cite{Khachatryan:2014vma} at $\sqrt{s}\,=$ 8 TeV using 19.7 fb$^{-1}$ of data. 
The measured helicity fractions are $F_{L}\,=\,0.298\,\pm\,0.028\,\mathrm{(stat)}\,\pm\,0.032\,\mathrm{(syst)}$, 
$F_{0}\,=\,0.720\,\pm\,0.039\,\mathrm{(stat)}\,\pm\,0.037\,\mathrm{(syst)}$ and 
$F_{R}\,=\,-0.018\,\pm\,0.019\,\mathrm{(stat)}\,\pm\,0.011\,\mathrm{(syst)}$. In addition limits are set on the real part 
of the tWb anomalous couplings, $g_{L}$ and $g_{R}$. This method is complementary to traditional measurements performed on 
$\mathrm{t\overline{t}}$ events~\cite{CMS-PAS-TOP-12-025}.

\section{Associated Production $t\overline{t}X$}
\label{sec:Associated Production}
\begin{figure}[ht!]
 \begin{minipage}{0.45\textwidth}
 \centering
 \includegraphics[width=\textwidth]{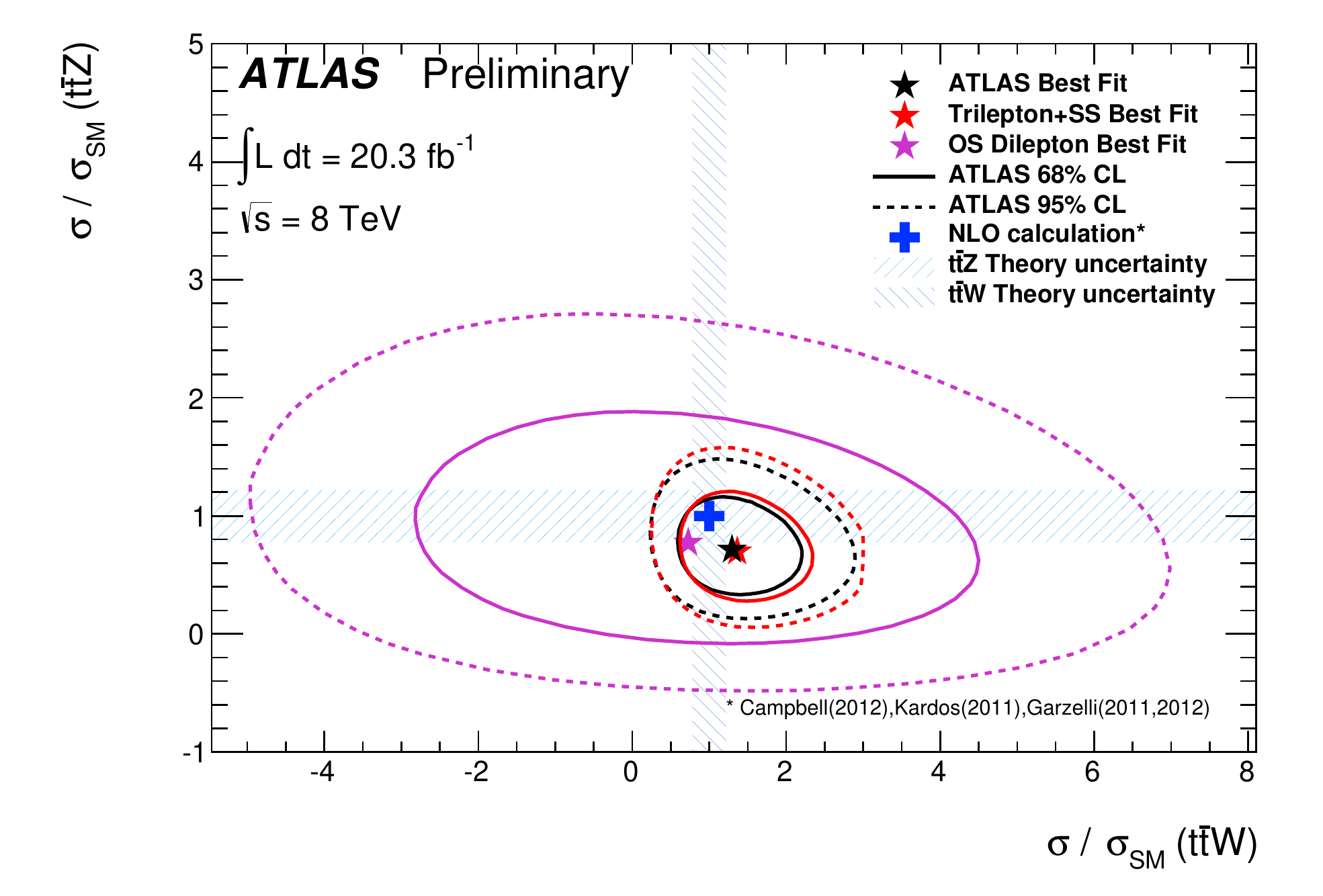} 
 \caption{$\sigma(t\overline{t}W)$ and $\sigma(t\overline{t}Z)$ cross sections compared to the SM values. The likelihood 
 contours are defined by the $\chi^{2}$ quantile for two degrees of freedom.~\protect\cite{ATLAS-CONF-2014-038} }
  \label{fig.ATLASTTX}
 \end{minipage}
 \hspace{0.2cm}
 \begin{minipage}{0.45\textwidth}
 \centering
 \includegraphics[width=\textwidth]{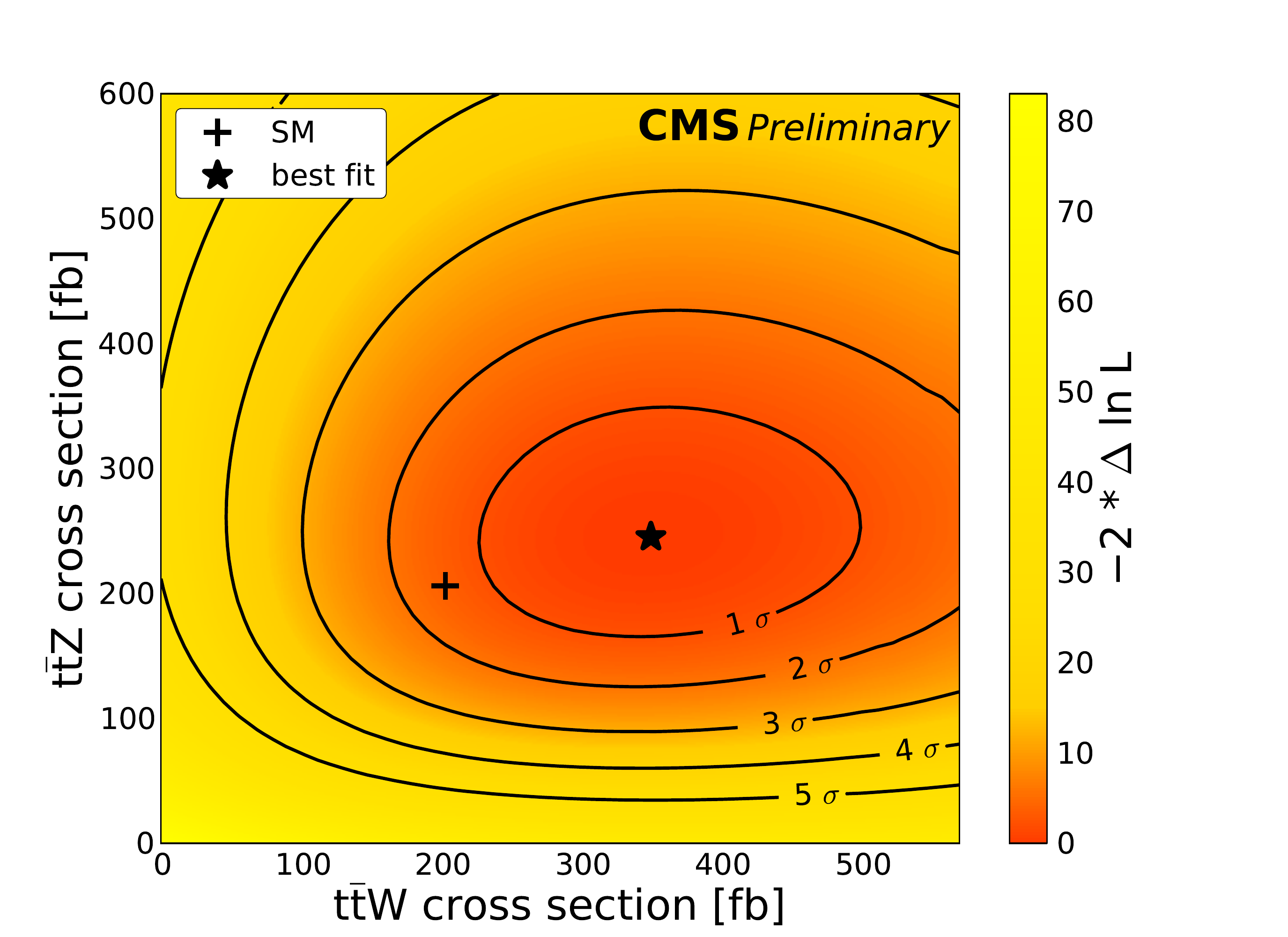}
 \caption{Profile likelihood as a function of $\sigma(t\overline{t}W)$ and $\sigma(t\overline{t}Z)$.~\protect\cite{TOP14021} }
  \label{fig.CMSTTX}
 \end{minipage}
 \end{figure}

The measurement of the cross section of vector boson production in association to $t\overline{t}$ pairs tests
the SM and offers a direct measurement of the coupling between the top quark and the Z, $\gamma$. These couplings can be used 
in searches for New Physics and 
constitute a background in $t\overline{t}H$. An observation of $t\overline{t}\gamma$ at $\sqrt{s}\,=$ 7 TeV is presented 
by the ATLAS collaboration~\cite{PhysRevD.91.072007} using 4.59 fb$^{-1}$ of data. The 
$t\overline{t}\gamma$ 
production cross section times the branching ratio (BR) of the lepton+jets decay channel is measured within the ATLAS detector 
and kinematic acceptance to be $\sigma^{fid}_{t\overline{t}\gamma} \times \mathrm{BR}\,=\,63\,\pm\,8\,\mathrm{(stat)}\,{}^{+17}_{-13}\,\mathrm{(syst)}\,\pm\,1\,\mathrm{(lumi)}\,\mathrm{fb}$. 
The result is in good agreement with the theoretical SM prediction of $48\,\pm\,10\,\mathrm{fb}$. The first measurement of the 
$t\overline{t}\gamma$ cross section at $\sqrt{s}\,=$ 8 TeV was performed by CMS~\cite{CMS-PAS-TOP-13-011} using 
19.7 fb$^{-1}$ of data. The inclusive $t\overline{t}\gamma$ production cross section for 
$p^{\gamma}_{T} > 20\, \mathrm{GeV}$ is measured as $\sigma_{t\overline{t}\gamma}\,=\,2.4\,\pm\,0.2\,\mathrm{(stat)}\,\pm\,0.6\,\mathrm{(syst)}\,\mathrm{pb}$ 
where the SM expectation is $\sigma^{SM}_{t\overline{t}\gamma}\,=\,1.8\,\pm\,0.5\,\mathrm{pb}$.\\
The associated production cross section of $t\overline{t}$ with either a W or a Z boson has been measured by both CMS and ATLAS 
at $\sqrt{s}\,=$ 8 TeV. Several signal categories are considered such as same-sign and opposite-sign dilepton, 
trilepton and fourlepton channels. In a simultaneous extraction of $\sigma_{t\overline{t}Z}$ and $\sigma_{t\overline{t}W}$, 
ATLAS~\cite{ATLAS-CONF-2014-038} provides evidence with $\sigma_{t\overline{t}Z}\,=\,150^{+55}_{-50}\,\mathrm{(stat)}\,\pm\,21\,\mathrm{(syst)}\,\mathrm{fb}$ 
and $\sigma_{t\overline{t}W}\,=\,300^{+120}_{-100}\,\mathrm{(stat)}\,{}^{+70}_{-40}\,\mathrm{(syst)}\,\mathrm{fb}$, both with 3.1 $\sigma$ 
observed excess over the background hypothesis. CMS~\cite{TOP14021} obtains a measurement of 
$\sigma_{t\overline{t}Z}\,=\,242^{+65}_{-55}\,\mathrm{fb}$ with 6.4 $\sigma$ observed excess over the background hypothesis 
and $\sigma_{t\overline{t}W}\,=\,382^{+117}_{-102}\,\mathrm{fb}$ with 4.8 $\sigma$ observed excess. The SM NLO predictions give 
$\sigma_{t\overline{t}Z}\,=\,206\,\pm\,29\,\mathrm{fb}$ and $\sigma_{t\overline{t}W}\,=\,203\,\pm\,25\,\mathrm{fb}$. These 
results of the simultaneous fits are shown in fig.~\ref{fig.ATLASTTX} for ATLAS and fig.~\ref{fig.CMSTTX} for CMS.

\section{Conclusions}
\label{sec:Conclusions}
Top quark properties measurements performed by the ATLAS and CMS collaborations at the LHC provide precision tests of the SM, 
also in single top quark production. All the results presented are consistent with the SM, with no indications 
of new physics yet. 

\section*{References}

\begin{thebibliography}{}

\end{thebibliography}


\begin{thebibliography}{99}
\bibitem{1748-0221-3-08-S08003}
ATLAS Collaboration,
\newblock JINST {\textbf 3}, S08003 (2008).

\bibitem{Chatrchyan:2008aa}
CMS Collaboration,
\newblock JINST {\textbf 3}, S08004 (2008).

\bibitem{PhysRevD.84.115013}
Aguilar-Saavedra J. A. and P\'erez-Victoria M.,
\newblock {\em Phys. Rev. D} \textbf {84}, 115013 (2011).

\bibitem{AguilarSaavedra:2011ug}
Aguilar-Saavedra J. A. and P\'erez-Victoria M.,
\newblock {\em JHEP} \textbf {09}, 097 (2011).

\bibitem{Aad:2013cea}
ATLAS Collaboration,
\newblock {\em JHEP} \textbf {02}, 107 (2014).

\bibitem{CMS-PAS-TOP-12-033}
CMS Collaboration,
\newblock CMS-PAS-TOP-12-033.

\bibitem{BernreutherAsym}
W. Bernreuther and Z.-G. Si,
\newblock {\em Phys. Rev.} D \textbf {86}, 034026 (2012).

\bibitem{ATLAS2015}
ATLAS Collaboration,
\newblock {\em JHEP} \textbf {05}, 061 (2015).

\bibitem{PhysRevLett.111.232002}
ATLAS Collaboration,
\newblock {\em Phys. Rev. Lett.} \textbf {111}, 232002 (2013).

\bibitem{PhysRevLett.112.182001}
CMS Collaboration,
\newblock {\em Phys. Rev. Lett.} \textbf {112}, 182001 (2014).

\bibitem{CMS-PAS-TOP-13-001}
CMS Collaboration,
\newblock CMS-PAS-TOP-13-001.

\bibitem{CMS-PAS-TOP-14-005}
CMS Collaboration,
\newblock CMS-PAS-TOP-14-005.

\bibitem{PhysRevLett.114.142001}
ATLAS Collaboration,
\newblock {\em Phys. Rev. Lett.} \textbf {114}, 142001 (2015).

\bibitem{Aad:2015pfx}
ATLAS Collaboration,
\newblock arXiv:1506.08616 (2015).

\bibitem{CMS-PAS-TOP-13-015}
CMS Collaboration,
\newblock CMS-PAS-TOP-13-015.

\bibitem{Khachatryan:2014vma}
CMS Collaboration,
\newblock {\em JHEP} \textbf{01}, 053 (2015).

\bibitem{CMS-PAS-TOP-12-025}
CMS Collaboration,
\newblock CMS-PAS-TOP-12-025.

\bibitem{PhysRevD.91.072007}
ATLAS Collaboration,
\newblock {\em Phys. Rev.} D \textbf{91}, 072007 (2015).

\bibitem{CMS-PAS-TOP-13-011}
CMS Collaboration,
\newblock CMS-PAS-TOP-13-011.

\bibitem{ATLAS-CONF-2014-038}
ATLAS Collaboration,
\newblock ATLAS-CONF-2014-038.

\bibitem{TOP14021}
CMS Collaboration,
\newblock CMS-PAS-TOP-14-021.

\end{thebibliography}

\end{document}